\documentclass[aps,prl,reprint,showpacs,superscriptaddress]{revtex4-1}

\usepackage{graphicx}
\usepackage{dcolumn}
\usepackage{bm}
\usepackage{color}
\usepackage{lipsum}
\usepackage{lineno}
\usepackage{soul}
\usepackage[colorlinks,urlcolor=blue,linkcolor=blue,citecolor=blue]{hyperref}

\usepackage{siunitx}

\begin{document}
	
\title{Observation of stopping power reduction at strong ion-plasma coupling}%
\author{Yun Liu}\thanks{These authors have contributed equally to this work.}
\affiliation{MOE Key Laboratory for Nonequilibrium Synthesis and Modulation of Condensed Matter, School of Physics, Xi’an Jiaotong University, Xi’an, 710049, China}

\author{Jieru Ren}\thanks{These authors have contributed equally to this work}\email{renjieru@xjtu.edu.cn}\affiliation{MOE Key Laboratory for Nonequilibrium Synthesis and Modulation of Condensed Matter, School of Physics, Xi’an Jiaotong University, Xi’an, 710049, China}

\author{Zhigang Deng}\thanks{These authors have contributed equally to this work}\email{dzgzju@163.com}
\affiliation{National Key Laboratory of Plasma Physics, Laser Fusion Research Center, China Academy of Engineering Physics, Mianyang, 621900, People’s Republic of China}

\author{Wei Qi}
\affiliation{National Key Laboratory of Plasma Physics, Laser Fusion Research Center, China Academy of Engineering Physics, Mianyang, 621900, People’s Republic of China}

\author{Bubo Ma}
\affiliation{MOE Key Laboratory for Nonequilibrium Synthesis and Modulation of Condensed Matter, School of Physics, Xi’an Jiaotong University, Xi’an, 710049, China}

\author{Wenqing Wei}
\affiliation{MOE Key Laboratory for Nonequilibrium Synthesis and Modulation of Condensed Matter, School of Physics, Xi’an Jiaotong University, Xi’an, 710049, China}

\author{Shizheng Zhang}
\affiliation{MOE Key Laboratory for Nonequilibrium Synthesis and Modulation of Condensed Matter, School of Physics, Xi’an Jiaotong University, Xi’an, 710049, China}

\author{Xuyang Luo}
\affiliation{MOE Key Laboratory for Nonequilibrium Synthesis and Modulation of Condensed Matter, School of Physics, Xi’an Jiaotong University, Xi’an, 710049, China}

\author{Ziqian Zhao}
\affiliation{MOE Key Laboratory for Nonequilibrium Synthesis and Modulation of Condensed Matter, School of Physics, Xi’an Jiaotong University, Xi’an, 710049, China}

\author{Mingzhe Yang}
\affiliation{MOE Key Laboratory for Nonequilibrium Synthesis and Modulation of Condensed Matter, School of Physics, Xi’an Jiaotong University, Xi’an, 710049, China}

\author{Yifang Gao}
\affiliation{MOE Key Laboratory for Nonequilibrium Synthesis and Modulation of Condensed Matter, School of Physics, Xi’an Jiaotong University, Xi’an, 710049, China}

\author{Xueguang Ren}
\affiliation{MOE Key Laboratory for Nonequilibrium Synthesis and Modulation of Condensed Matter, School of Physics, Xi’an Jiaotong University, Xi’an, 710049, China}

\author{Jianxing Li}
\affiliation{MOE Key Laboratory for Nonequilibrium Synthesis and Modulation of Condensed Matter, School of Physics, Xi’an Jiaotong University, Xi’an, 710049, China}

\author{Dieter H. H. Hoffmann}
\affiliation{MOE Key Laboratory for Nonequilibrium Synthesis and Modulation of Condensed Matter, School of Physics, Xi’an Jiaotong University, Xi’an, 710049, China}

\author{Xing Wang}
\affiliation{MOE Key Laboratory for Nonequilibrium Synthesis and Modulation of Condensed Matter, School of Physics, Xi’an Jiaotong University, Xi’an, 710049, China}

\author{Zhongfeng Xu}
\affiliation{MOE Key Laboratory for Nonequilibrium Synthesis and Modulation of Condensed Matter, School of Physics, Xi’an Jiaotong University, Xi’an, 710049, China}

\author{Shaoyi Wang}
\affiliation{National Key Laboratory of Plasma Physics, Laser Fusion Research Center, China Academy of Engineering Physics, Mianyang, 621900, People’s Republic of China}

\author{Quanping Fan}
\affiliation{National Key Laboratory of Plasma Physics, Laser Fusion Research Center, China Academy of Engineering Physics, Mianyang, 621900, People’s Republic of China}

\author{Bo Cui}
\affiliation{National Key Laboratory of Plasma Physics, Laser Fusion Research Center, China Academy of Engineering Physics, Mianyang, 621900, People’s Republic of China}

\author{Weiwu Wang}
\affiliation{National Key Laboratory of Plasma Physics, Laser Fusion Research Center, China Academy of Engineering Physics, Mianyang, 621900, People’s Republic of China}

\author{Sixin Wu}
\affiliation{National Key Laboratory of Plasma Physics, Laser Fusion Research Center, China Academy of Engineering Physics, Mianyang, 621900, People’s Republic of China}

\author{Yue Yang}
\affiliation{National Key Laboratory of Plasma Physics, Laser Fusion Research Center, China Academy of Engineering Physics, Mianyang, 621900, People’s Republic of China}

\author{Zhurong Cao}
\affiliation{National Key Laboratory of Plasma Physics, Laser Fusion Research Center, China Academy of Engineering Physics, Mianyang, 621900, People’s Republic of China}

\author{Zongqing Zhao}
\affiliation{National Key Laboratory of Plasma Physics, Laser Fusion Research Center, China Academy of Engineering Physics, Mianyang, 621900, People’s Republic of China}

\author{Yuqiu Gu}
\affiliation{National Key Laboratory of Plasma Physics, Laser Fusion Research Center, China Academy of Engineering Physics, Mianyang, 621900, People’s Republic of China}

\author{Leifeng Cao}
\affiliation{Advanced Materials Testing Technology Research Center, Shenzhen Technology University, Shenzhen, 518118, China}

\author{Bin He}
\affiliation{Institute of Applied Physics and Computational Mathematics, Beijing 100094, China}

\author{Shaoping Zhu}
\affiliation{Institute of Applied Physics and Computational Mathematics, Beijing 100094, China}
\affiliation{National Key Laboratory of Plasma Physics, Laser Fusion Research Center, China Academy of Engineering Physics, Mianyang, 621900, People’s Republic of China}
\affiliation{Graduate School, China Academy of Engineering Physics, Beijing 100088, China} 

\author{Olga Rosmej}
\affiliation{Helmholtzzentrum für Schwerionenforschung GSI, 64291 Darmstadt, Germany}
\affiliation{Goethe-University Frankfurt, Institute of Applied Physics, 60438 Frankfurt am Main, Germany}

\author{Rui Cheng}
\affiliation{Institute of Modern Physics, Chinese Academy of Sciences, Lanzhou 710049, China}

\author{Guoqing Xiao}
\affiliation{Institute of Modern Physics, Chinese Academy of Sciences, Lanzhou 710049, China}

\author{Weimin Zhou}
\email{zhouwm@caep.cn}
\affiliation{National Key Laboratory of Plasma Physics, Laser Fusion Research Center, China Academy of Engineering Physics, Mianyang, 621900, People’s Republic of China}

\author{Yongtao Zhao}
\email{zhaoyongtao@xjtu.edu.cn}
\affiliation{MOE Key Laboratory for Nonequilibrium Synthesis and Modulation of Condensed Matter, School of Physics, Xi’an Jiaotong University, Xi’an, 710049, China}

	\bibliographystyle{apsrev4-1}

\date{\today}
	
\begin{abstract}
{Ion stopping in dense plasma is crucial for stellar evolution and fusion ignition. However, its behavior in the strong ion-plasma coupling regime beyond the linear limit has long remained elusive, due to formidable experimental challenges. Here we report the first experimental investigation of ion stopping at an unprecedented coupling parameter exceeding unity, achieved by sending laser-accelerated short-pulse and intense quasi-monoenergetic carbon ions ($\sim$583 keV/u, C$^{5+}$) into a uniform, long-lived, well-characterized dense plasma target ($T_e$ $\approx$ 17 eV, $n_e$ $\approx$ 4$\times$10$^{20}$ cm$^{-3}$). By simultaneously measuring ion energy loss and charge-state evolution, we eliminated key experimental ambiguities arising from charge-state determination. Our results clearly show a reduction in stopping power compared with predictions from standard linear dielectric response or binary collision models, and they agree well with the hybrid calculation of molecular dynamics with quantum corrections. The importance of nonlinear screening effects arising from many-body interactions and quantum effects due to the wave nature of electrons was demonstrated at strong coupling. This work establishes a definitive high-fidelity experimental benchmark for collisional dynamics in the strong-coupling regime. It offers critical insight for accurate modeling of energy transport in inertial confinement fusion and astrophysical plasmas.}
\end{abstract}                        
	\maketitle

A projectile moving through a dense plasma experiences a drag force-stopping power dE/dx-that determines how quickly it transfers energy to its surroundings. This is why alpha particles from fusion reactions either ignite the fuel or escape \cite{labaune2013fusion,zylstra2022burning,williams2024demonstration}, why ion beams either heat matter or pass through it unperturbed, why fast ions in accretion disks either power X-ray emission by heating the plasma or carry their energy away \cite{Remington1999,Frank2002}. Besides, it provides a velocity-resolved probe of the collisional dynamics and many-body interactions that control the thermally averaged plasma transport properties, such as conductivity \cite{Lambert2011,Starrett2012,Wang2012,Faussurier2015,Hu2016} and temperature relaxation \cite{Dimonte2008,Ma2019,Rightley2021,Kodanova2025}.  Hence, the ion-stopping
process has attracted broad interest in the fields of inertial confinement fusion \cite{Atzeni2004,Hurricane2014,Betti2015,Betti2016,Sio2019fuel,Hartouni2023,Hurricane2024}, ion-beam-driven high-energy-density physics \cite{Roth2001,Wang2015,Kawata2016,Hofmann2018}, astrophysics \cite{Patel2003,Duffy2008,Sharkov2016high,Casey2017,Ren2017NIMB} and extreme states of matter diagnostics \cite{Mackinnon2006,Volpe2011,Zylstra2015,Xu2017,Sun2017,hayes2020plasma,Schoenberg2020}. Yet, despite its ubiquity and importance, a long-standing discrepancy persists in describing this process in regimes of strong ion-plasma coupling, where the projectile strongly perturbs the surrounding electron cloud, leading to the failure of linear theoretical frameworks.

\begin{figure*}[t!]
	\centering
	\includegraphics[width=\linewidth]{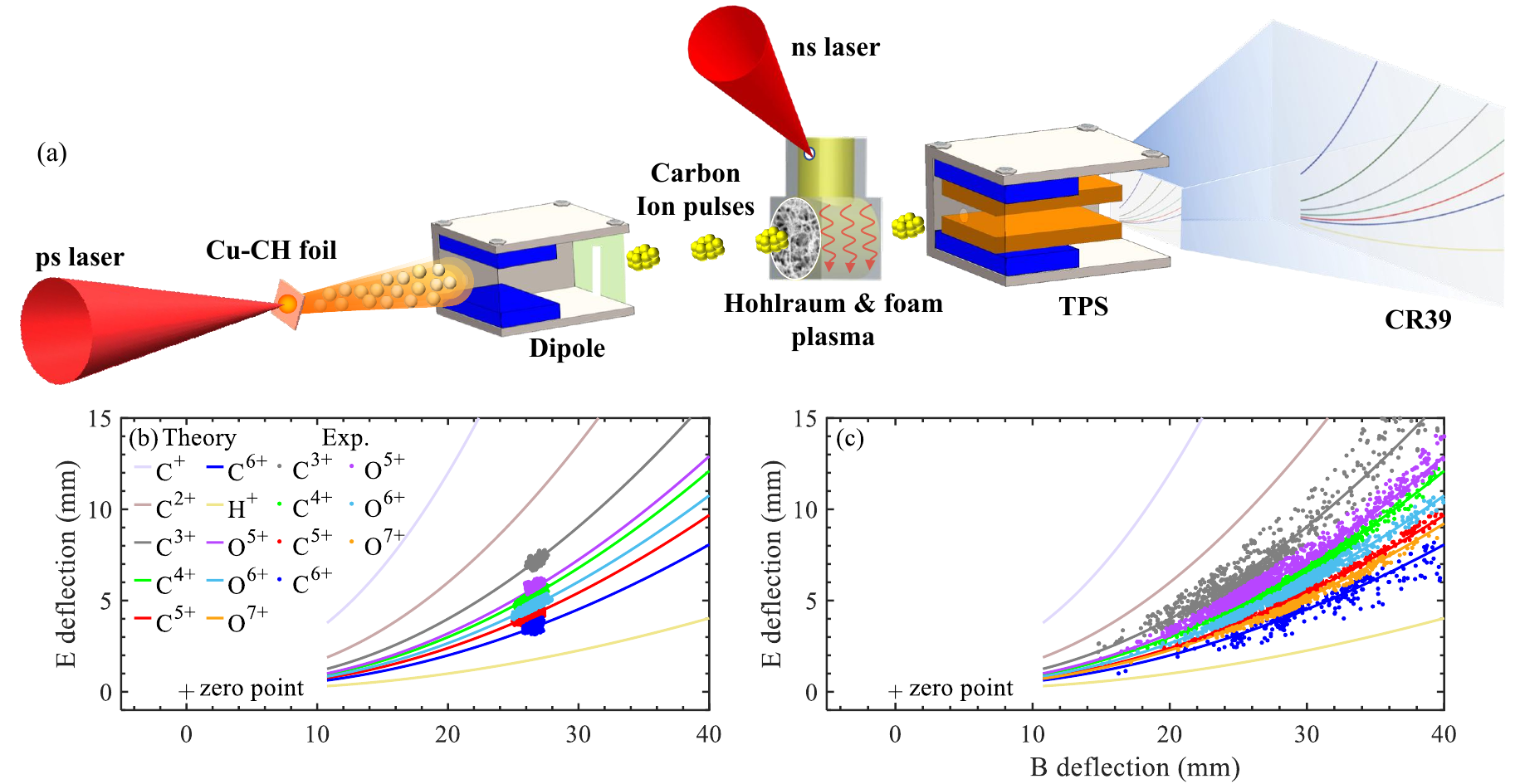}
	\caption{Layout of the experimental setup. (a) A picosecond laser is focused onto a CH-coated copper foil, generating an intense short-pulse ion beam with broad energy spectrum via the TNSA mechanism. A magnetic dipole with entrance and exit slits trims out the quasi-monoenergetic ions with the same momentum-to-charge ratio. After some flying distance, ions are naturally separated into pulses according to their species and charge state and interact with the foam target, which was heated by the ns laser-driven hohlraum radiation to generate dense plasma. The ions passing through the plasma are detected by a TPS coupled to CR39 film. (b) Thomson parabola tracks of laser-accelerated quasi-monoenergetic carbon and oxygen ions recorded on a CR39 detector without target, along with theoretical deflection curves. (c) Thomson parabola tracks of carbon and oxygen ions passing through target, as well as the theoretical deflection curves. In (b) and (c), the colored dots represent the experimentally-recorded ion tracks, the '+' symbol marks the zero-reference point, and solid curves indicate the theoretical deflection distances for different ion species.}
	\label{fig:figure1}
\end{figure*}

According to Thomas Peter and Meyer-ter-Vehn, the ion-plasma coupling strength can be quantified approximately by the ratio of effective ion charge $Z_{eff}$ to the
electron number in the Debye sphere, namely $\mathcal{Z}\ =\frac{Z_{eff}}{n_0\lambda_D^3}$ \cite{Peter1991}, where $n_0$ and $\lambda_D$ are the electron density and the Debye length, respectively. In the weak-coupling limit ($\mathcal{Z}$ $\ll$ 1), the plasma's dielectric response is linearly proportional to the ion's charge, leading to the well-known scaling dE/dx $\propto$ $Z_{eff}^2$. Bethe’s classic quantum-mechanical formulation \cite{Bethe1930,Bloch1933}, standard stopping model (SSM) \cite{Deutsch2016Ion}, modern Li-Petrasso (LP) \cite{LP1993} and Brown-Preston-Singleton (BPS) approaches \cite{Brown2005} have been applied. Nevertheless, for slow and/or highly charged ions in dense plasmas close to the end of their range, the ion-plasma coupling could be strong enough to trigger a transition to a nonlinear coupling regime \cite{Peter1991,Zwicknagel1999,Zwicknagel2002}, where conditions for the linearized treatment are violated. The theoretical predictions, even regarding the basic scaling with ion charge, diverge starkly \cite{Peter1991,Zwicknagel2002}. The lack of high-fidelity experimental benchmarks has left the field without a consensus.

Experimental progress on ion stopping in dense plasma has long been hindered by the difficulty of simultaneously generating and combining intense quasi-monoenergetic ions with a well-defined dense plasma whose hydrodynamic evolution timescale exceeds the ion transit time, thereby enabling reasonable precision. Based on high-power lasers and accelerators, experiments on ion stopping in dense plasmas have recently been reported \cite{jacoby1995stopping,Frenje2015,Zylstra2015,Cayzac2017,Sayre2019,Frenje2019,Malko2022,Frank2013,Braenzel2018}, with parameters remaining largely within the linear regime. Experimental validation in the nonlinear ion-plasma coupling regime has remained virtually unexplored, creating a critical gap between theory and observation. 
	
In this letter, we directly probe stopping power in an unprecedented nonlinear ion-plasma coupling regime ($\mathcal{Z}$ > 1) using a novel experimental platform. We send a laser-accelerated, intense, quasi-monoenergetic (583 keV/u) short-pulse C$^{5+}$ ion beam into a long-lived, homogeneous, well-defined plasma target ($T_e$ $\sim$ 17 eV, $n_e$ $\sim$ 4$\times$10$^{20}$ cm$^{-3}$) generated by heating a foam target with hohlraum X-ray radiation. The beam-target interaction timescale of $\sim$ 0.26 ns is much shorter than the typical plasma evolution timescale of $\sim$ 10 ns, excluding uncertainties arising from vagueness in beam and target parameters. The single-shot, simultaneous measurement of ion energy loss and the charge-state distribution as they evolve through the plasma eliminates the uncertainty associated with the assumed effective charge $Z_{eff}$. These capabilities enable testing of state-of-the-art plasma stopping-power theories with sufficient precision. Our measurements clearly reveal a significant reduction in stopping power compared to predictions from all standard weak-perturbation or linear-response models. This reduction is closely reproduced by classical molecular dynamics simulations that self-consistently capture non-linear screening, but only when they are augmented with a quantum correction. This work establishes a precise experimental benchmark and a platform for probing complex collisional dynamics in plasma.

The experiment was performed at the XG-III laser facility of the Laser Fusion Research Center in Mianyang, using a dual-beam configuration. The experimental layout is displayed in Fig. \ref{fig:figure1}(a). A high-power laser beam of 131 J of total energy, 759 fs duration, and a 20 $\mu$m focal spot was focused on a 15 $\mu$m-thick CH-coated copper foil to generate ions via the target normal sheath acceleration (TNSA). These ions typically had a broad energy distribution, which is unfavorable for energy-loss analysis. To address this limitation, a 0.28-Tesla magnetic dipole was used to trim the quasi-monoenergetic ion beams. A 500 $\mu$m entrance slit first collimated the ions, which were then laterally dispersed by the dipole. After passing through a 500 $\mu$m exit slit, the quasi-monoenergetic ion pulses with the same momentum-to-charge ratio were selected and hit the target pulse-by-pulse, with time delays according to their velocities.

The target consists of a gold hohlraum converter and a porous tricellulose-acetate (TCA, C$_9$H$_{16}$O$_8$) foam (2 mg/cm$^3$ density, 1 mm thickness). The laser beam with a total energy of 147 J and a 2 ns duration irradiates the inner surface of the hohlraum to produce soft X-rays, which subsequently heat the foam to a plasma state. This heating scheme, which has a great advantage for generating
homogeneous, 10 ns-long-lived plasmas, has been extensively studied at the Phelix \cite{rosmej2011heating,faik2014creation,rosmej2015hydrodynamic} and XGIII laser facilities \cite{Ren2020,Ren2023,Ma2021,Ma2024,Cheng2025}. Diagnostics confirmed a plasma temperature of 17$\pm$1 eV and a free electron density of (4.0$\pm$0.3)$\times$10$^{20}$ cm$^{-3}$. This partially ionized plasma is still in the ideal regime with electron coupling of $\Gamma_{ee} = \frac{e^2}{a_ek_BT}$ $\sim$ 0.09 and degeneracy of $\Theta$ = $\frac{k_BT}{E_F}$ $\sim$ 100, where $e$ is the elementary charge, $a_e$ = $(\frac{3}{4 \pi n_e})^{1/3}$ is average distance between the electrons, $k_B$ is Boltzmann's constant, $T$ is the temperature, and $E_F$ is the Fermi energy.
	
	\begin{figure}[ht!]
		\centering
		\includegraphics[width=1\linewidth]{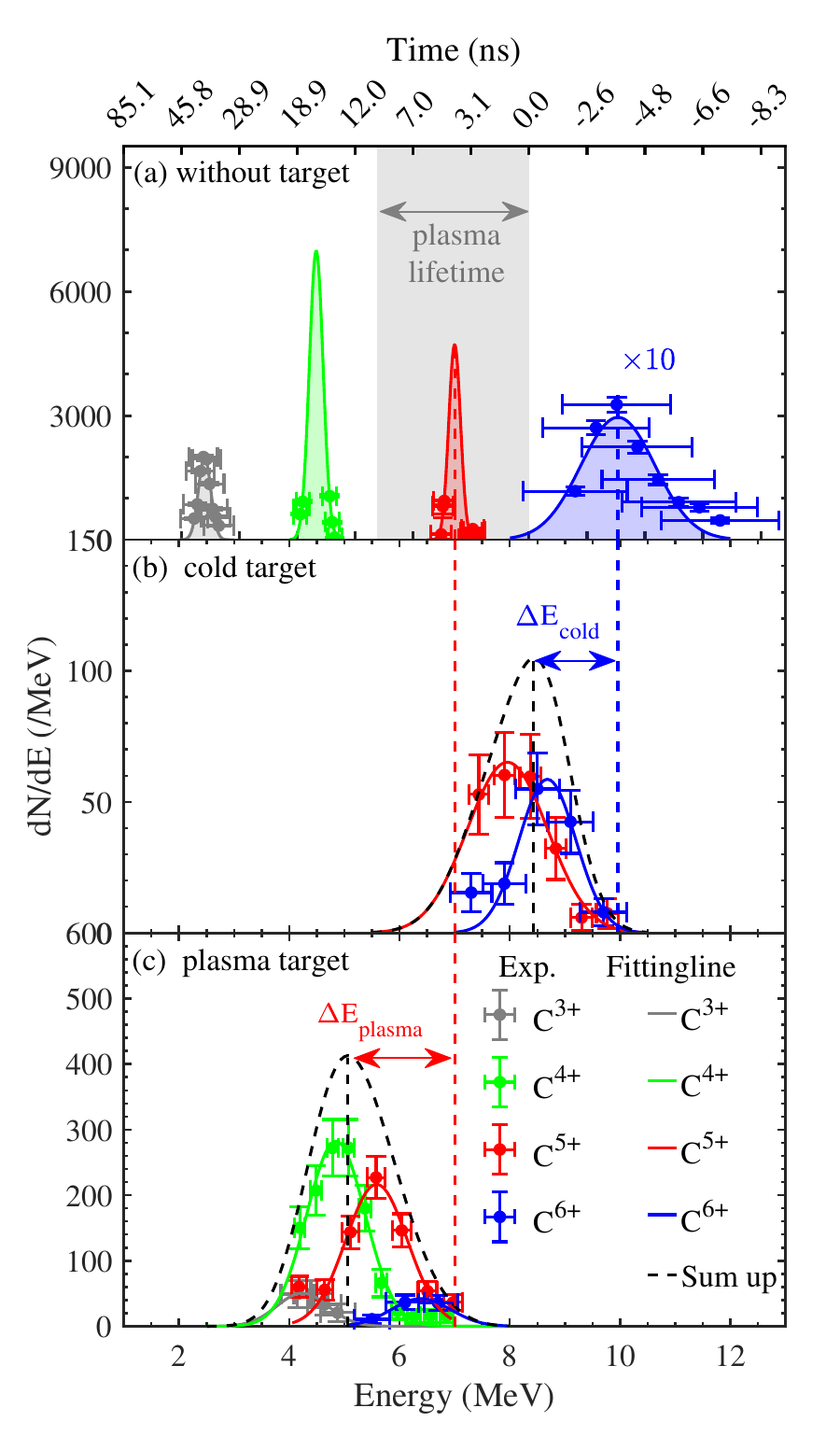} 
		\caption{Energy spectra of the trimmed quasi-monoenergetic carbon ions without any target (incident reference), as well as ions passing through unheated (cold) and heated (plasma) target. (a) Energy spectra of incident quasi-monoenergetic carbon ion pulses. The top axis indicates the relative arrival time of the ions at the foam target, with time zero defined as the ns laser trigger. The count of C$^{6+}$ ions has been multiplied by 10 for better visibility. (b) Energy spectra of ions that passed through cold foam, originating from quasi-monoenergetic C$^{6+}$ ions incidence. (c) Energy spectra of ions that passed through plasma target, originating from quasi-monoenergetic C$^{5+}$ ions incidence. In all panels, the experimental data are fitted with Gaussian profiles to guide the eyes. The energy loss was determined by the central peak energy downshift.}                                                       
		\label{fig:figure2}
	\end{figure}
	
The energy and charge-state distribution of the ion beam were recorded using a Thomson parabola spectrometer (TPS) coupled to CR39 track detector. The measured tracks of the heavy ions, excluding the protons, are shown by dots in Fig. \ref{fig:figure1}(b) and (c)  for the cases without and with the target, respectively. The X and Y coordinates represent the magnetic and electric deflection distances relative to the zero order. The solid curves represent the theoretical deflection distance of carbon and oxygen ions species. The accelerated heavy ions consist of carbon ions (C$^{3+}$, C$^{4+}$, C$^{5+}$, and C$^{6+}$) and oxygen ions (mainly O$^{5+}$ and O$^{6+}$). In this manuscript, only carbon-ion stopping is discussed.

The deflection distances of carbon ions without and with the target are converted to energies in Fig. \ref{fig:figure2}(a) and Fig. \ref{fig:figure2}(b-c), respectively, where vertical error bars indicate statistical uncertainties and horizontal error bars represent the energy resolution of TPS. To guide the eyes, the energy spectra were fitted by Gaussian profiles. For the C$^{5+}$ and C$^{4+}$, the CR39 detector is saturated near the central energy region, so the unsaturated tracks in the high- and low-energy tails are counted in the fitting and central energy determination. The central energies of the selected initial C$^{3+}$, C$^{4+}$, C$^{5+}$, and C$^{6+}$ ions without plasma target are 2.47 MeV, 4.49 MeV, 7.00 MeV, and 10.01 MeV, respectively, which agrees well with the rule of the same momentum-to-charge ratio. 
	
The top axis of Fig. \ref{fig:figure2} shows the arrival time of carbon ions at the target, referenced to the laser trigger time (t = 0). The shaded region marks the 10 ns window during which the foam target is heated but has not yet expanded significantly. Within such time sequence, the C$^{6+}$ ions, which arrive earlier, interact with the cold foam, whereas the C$^{5+}$ ions which arrive later, interact with the uniform, well-characterized plasma target. The ion-plasma interaction timescale for C$^{5+}$ is approximately 0.26 ns, which is significantly shorter than the plasma evolution timescale ($\sim$ 10 ns), indicating that the plasma target is quasi-static. By the time the slower 4.49 MeV C$^{4+}$ and 2.47 MeV C$^{3+}$ ions arrive at the target, the plasma has already expanded significantly, and its state is not well known. Therefore, we discuss only the cases in which C$^{6+}$ and C$^{5+}$ traverse the target.
	
The energy distributions for all the carbon ions with energies above 4 MeV after the target are shown in Fig. \ref{fig:figure2}(b-c). The ions shown in Fig. \ref{fig:figure2}(b) are expected to originate from the C$^{6+}$ projectile interacting with the cold foam because all other carbon-ion projectiles initially have lower energies and have little chance of gaining energy after passing through the target. It can be seen that after passing through the cold foam, the ion species shifted from C$^{6+}$ to C$^{5+}$ and C$^{6+}$ due to charge-exchange processes, and the central energy of the overall ion distribution was downshifted from 9.96 MeV to 8.45 MeV, indicating an energy loss of 1.51 MeV in the cold foam. This energy loss agrees well with the well-established SRIM code \cite{ziegler1985stopping}, which validated our measurement. The ions shown in Fig. \ref{fig:figure2}(c) are supposed to originate from the C$^{5+}$-ions interaction with plasma, because C$^{4+}$-ions and C$^{3+}$-ions, after passing through the target, will have their energy down-shifted to a much lower energy region. After passing through the plasma target, the ion species of C$^{5+}$ spreads to C$^{3+}$, C$^{4+}$, C$^{5+}$, and C$^{6+}$, and the central energy of the overall ion distribution was downshifted from 7.00 MeV to 5.08 MeV, indicating an energy loss of 1.92 MeV.

\begin{figure}[t!]
	\centering
	\includegraphics[width=1\linewidth]{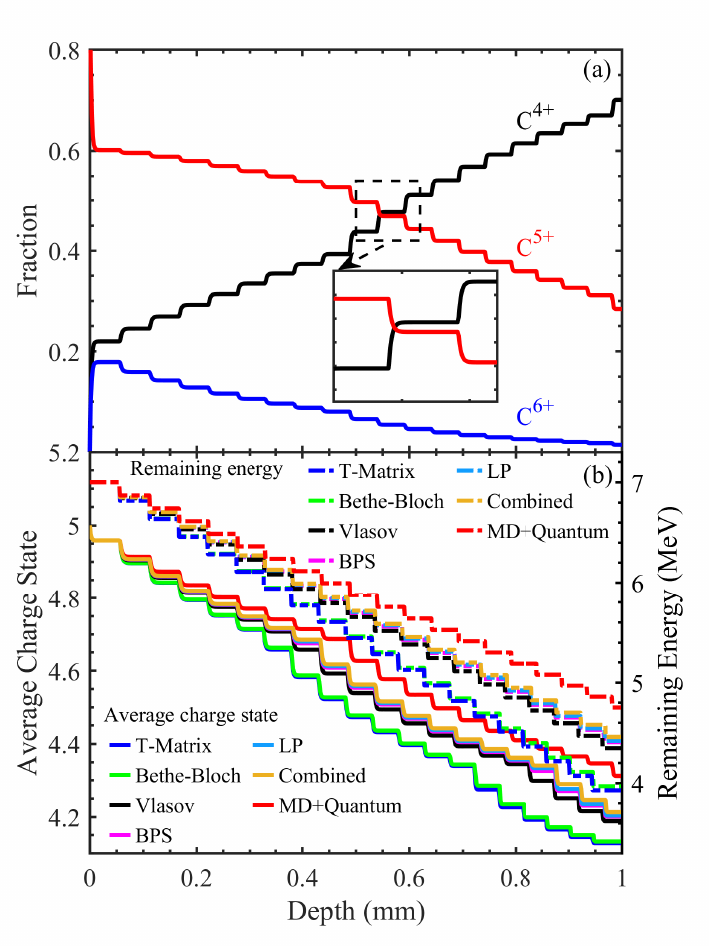}
	\caption{Hybrid charge state evolution and stopping simulations along the ion penetration depth in the plasma. (a) Typical carbon ion charge fraction evolution in the plasma target when atomic-state population kinetics and MD stopping theories are considered. Fractions of C$^{1+}$, C$^{2+}$ and C$^{3+}$ are 7 orders of magnitude lower, and hence are omitted for clarity. The inset implies the beam reaches charge equilibrium within a very short distance ($\sim$ 0.01 mm) in each calculation step. (b) Average charge state and residual energy evolution employing different stopping theories.}
	\label{fig:figure3}
\end{figure}

To resolve discrepancies among ion-stopping formalisms routinely used in the inertial confinement fusion community and to capture the main physical processes in the nonlinear, strongly coupled ion-plasma regime, we performed hybrid atomic-state population-kinetics and stopping simulations that incorporate different stopping models. In the simulation, once 7.00 MeV C$^{5+}$ ions enter the plasma, they undergo complex charge-changing processes during collisions with bound and free electrons. These ions soon evolve into a broad charge-state distribution. This distribution determined the effective charge state $Z_{eff}$ and the $Z_{eff}$-dependent energy loss dE/dx according to various stopping theories. The reduction in velocity due to energy loss changes the charge-changing cross sections and alters the following charge-state distribution as well as stopping power at deeper penetration depth, which forms one calculation loop. In our case, the total 20 loops were carried out until the ions traversed the 1 mm-thick target. The details are presented in Supplemental Material.

In our charge-state distribution calculations, rate equations are solved accounting for the dominant charge transfer processes, including Coulomb ionization by both bound and free electrons, capture of bound and free electrons, and three-body recombination. We also included the contributions from excited states via target-density-effect modifications \cite{rosmej2011heating,Rosmej_Excited_2007}, which have been experimentally shown to play an important role in the current plasma regime \cite{Ren2023,Ma2024,Cheng2025}. In the energy loss calculations, the commonly used formulas like Bethe-Bloch \cite{Bethe1930,Bloch1933}, T-Matrix \cite{Gericke2002}, Vlasov \cite{Peter1991}, LP \cite{LP1993} and BPS \cite{Brown2005} formalisms, combined models proposed by Zwicknagel \cite{Zwicknagel1999,Zwicknagel2002}, and Molecular Dynamic (MD) approaches \cite{Grabowski2013molecular} are used at every time step. In all cases, the energy deposition resulting from bound electrons is calculated with the corresponding terms of the Bethe-Bloch formula, which explicitly separates contributions from bound and free electrons.

\begin{figure}[t!]
	\centering
	\includegraphics[width=1\linewidth]{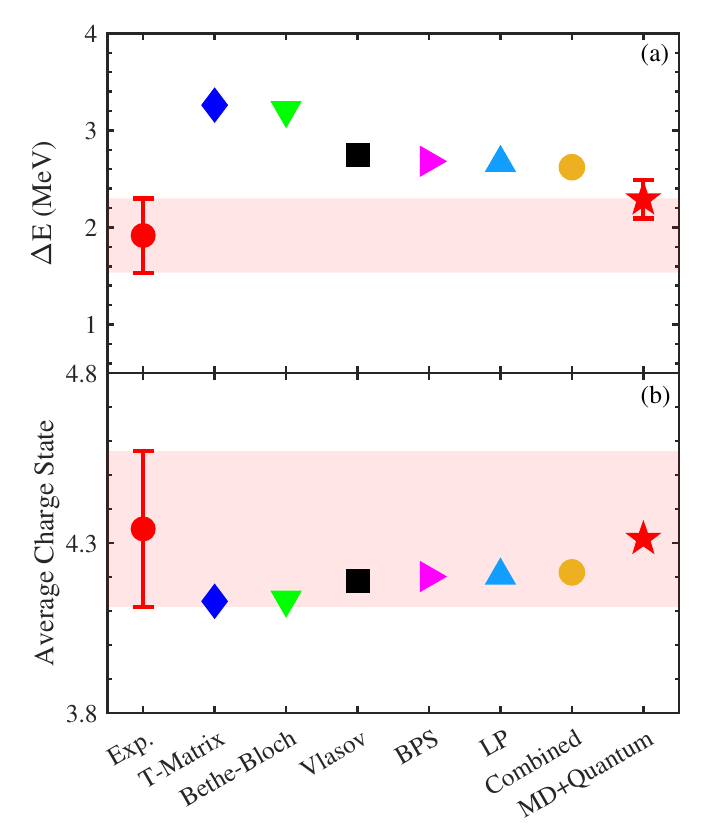}
	\caption{Comparison of experimentally measured energy loss and average charge state with hybrid simulation results employing different stopping models for ions passing through the plasma. The red dots represent experimental measured value. The stopping models including T-Matrix, Bethe-Bloch, Vlasov, BPS, LP, combined model proposed by Zwicknagel and classical Molecular Dynamics (MD) results with quantum correction, are used. (a) Measured energy loss versus simulation results. The error bar for the experimental data originates from the statistical error, fitting error, and the energy resolution of the TPS. For the simulation result that uses MD stopping theory and quantum correction (red star), the error bar reflects the spread among different quantum correction models. (b) Measured average charge state of ions passing through the plasma versus simulation results with the same color scheme as in (a).}
	\label{fig:figure4}
\end{figure}

The typical evolution of ion charge-state fractions, calculated in combination with MD stopping model, is shown in Fig. \ref{fig:figure3}(a), where the fractions of C$^{1+}$, C$^{2+}$, and C$^{3+}$ are about 7 orders of magnitude lower and therefore not displayed here. It can be seen that the ion beam reaches equilibrium charge within a very short distance ($\sim$ 0.01 mm) in every calculation step, because the characteristic charge-exchange timescale is much shorter than the stopping time \cite{Peter1991II}. Consequently, the effective charge along the ion trajectory $x$ depends instantaneously on the local ion velocity $Z_{eff}\left(x\right)=Z_{eq}(v(x))$. Fig. \ref{fig:figure3}(b) presents the corresponding evolution of the average charge state and the residual ion energy as a function of penetration depth when different stopping theories (T-Matrix, Bethe-Bloch, Vlasov, LP, BPS, Combined and MD) are applied. Theories that predict stronger energy loss yield lower average charge states, because lower ion velocities favor recombination over ionization. 

The measured energy loss and average charge state of ions passing through the plasma are compared with theoretical predictions in Fig. \ref{fig:figure4}(a-b). Calculations employing widely used stopping models, such as T-Matrix, Bethe-Bloch, Vlasov, BPS, and LP, systematically overestimate the energy loss and consequently underestimate the average charge state. These models are based on two-body collision or linear response approximations and are therefore primarily valid in the weak ion-plasma coupling regime. In the current experiment, if we take 6 MeV C$^{5+}$ ions as a typical example-this 6 MeV energy is the average of incident and exiting ions, and the charge state 5+ approximates the effective charge at 6 MeV. According to the Meyer-ter-Vehn definition, the ion plasma coupling parameter is $\mathcal{Z}$ $\approx$ 3.8, exceeding unity, and thus this beam-plasma interaction system enters the strong-coupling regime. Actually, BPS theory and Zwicknagel also defined the ion-plasma coupling strength. For example, BPS theory requires the Coulomb coupling parameter between projectile and plasma target $g$ $\ll$ 1 \cite{Brown2005}
\begin{equation}g^2=\sum_b\beta_b^2\left(\frac{Z_{eff}Z_b}{4\pi}\right)^2\kappa_b^2\end{equation}
where $Z_{eff}$ and $Z_b$ denote the charges of the projectile and plasma species $b$, $\kappa_b=\sqrt{\beta_be_b^2n_b}$ is the Debye wave number of species $b$, $n_b$ is the number density of species $b$, and $\beta_b=\frac{1}{T_b}$ ($T_b$ indicating the temperature of species $b$). Alternatively, Zwicknagel et al. identified the weak coupling by the condition $\left\langle \eta \right\rangle$ $\ll$ 1 \cite{Zwicknagel2002}
\begin{equation}\left\langle \eta \right\rangle = \frac{Z_{eff}e^2}{4\pi\varepsilon_0\hbar \left\langle v_r \right\rangle}\end{equation}
where $e$ is the elementary charge, $\varepsilon_0$ is the vacuum permittivity, $\hbar$ is the reduced Planck constant, and $\left\langle v_r \right\rangle$ is the averaged relative velocity between the projectile ion and plasma electrons. In our case, the values $\mathcal{Z}$ $\sim$ 3.8, $g$ $\sim$ 2.1, and $\eta$ $\sim$ 1.1 indicate that our system is in the strong-coupling regime, leading to the failure of these two-body collision or linear-response theories.

To describe the stopping process in the non-linear coupling regime. Zwicknagel proposed a combined approach \cite{Zwicknagel1999,Zwicknagel2002} that augments linear-response theory with corrections that incorporate contributions from the strong-coupling region induced by the ion. While prediction from this model moves slightly closer to the measured stopping power, it remains substantially higher. This discrepancy arises because the method still assumes that static and dynamic collective screening are well described by linear response-an approximation that breaks down once nonlinear screening becomes significant. As Zwicknagel reported later, the combined model scheme is therefore applicable only in regimes where nonlinear screening effects remain negligible \cite{Zwicknagel2002}. 

MD simulations provide a first-principles, non-perturbative framework capable of capturing the inherently nonlinear screening and many-body interactions in strongly coupled plasmas. The distance of closest approach is comparable to the electron de Broglie wavelength; hence, the wave nature of the electrons tends to soften the effective interaction potential between charged particles and thus reduces the stopping power \cite{Brown2005,Zwicknagel2002,Malko2022}.
In this work, we performed hybrid calculations based on the classical MD simulation result reported by Grabowski et al. \cite{Grabowski2013molecular} augmented with quantum corrections. We evaluate this quantum correction using two independent methods proposed by Brown et al. \cite{Brown2005} and Issanova et al. \cite{Issanova2016}, respectively. Since the two approaches yield different results, we adopt their average and assign an uncertainty that reflects the deviation between them. The resulting quantum-corrected MD prediction for the stopping power agrees with the measured energy loss within the uncertainties. It more accurately reproduces the observed average charge states in Fig. \ref{fig:figure4}(b).

In summary, the ion stopping and charge state evolution in a dense plasma under unprecedented, nonlinear, strongly coupled ion-plasma interaction conditions have been accurately measured. By using a short-pulse($\sim$ 0.26 ns), well-characterized, quasi-monoenergetic ion beam (7.0 MeV C$^{5+}$) and a long-lived ($\sim$ 10 ns), homogeneous plasma target ($T_e$ $\approx$ 17 eV, $n_e$ $\approx$ 4$\times$10$^{20}$ cm$^{-3}$), the routinely used stopping models are tested excluding the uncertainties from and beam-target parameter and charge state evolution histories. Our experimental data reveal a clear reduction in energy loss compared to predictions from established binary-collision and weak-perturbative theories such as Bethe-Bloch, BPS, Vlasov, T-Matrix, and LP et al. Agreement with the hybrid calculations combining classical molecular dynamics with quantum correction demonstrated the importance of (i) nonlinear screening arising from many-body interactions, which modified the ion-induced shielding cloud and potential beyond the linear mean field approximation due to the strong disturbance, and (ii) quantum effects that soften the interaction potential due to the wave nature of electrons. The synergistic action of these effects underpins the observed reduction in stopping power. This understanding is essential for predictive modeling in inertial confinement fusion (e.g., $\alpha$-particle energy deposition) and for accurately describing the energy transport in astrophysical environments such as stellar and accretion disks. Future efforts should extend this framework to broader parameter spaces, especially for slower, highly charged ions in dense plasma, where the coupling effects are supposed to be more pronounced. We want to point out that advanced first-principles calculations that incorporate non-linear screening and quantum effects in a self-consistent manner would be highly desirable.

The experiment was performed at the XG-III facility in Mianyang. The authors are grateful to the staff of Laser Fusion Research Center. The work was supported by the Chinese Science Challenge Project No. TZ2025012, the National Key R\&D Program of China, No. 2022YFA1603300, the National Natural Science Foundation of China (Grants No. 12422512, No. 12595365, No. U2541245, No. 12120101005,  No. 12405238, No. 12175174, No. 12325406, and No. 92261201), the Postdoctoral Research Funding Project of Shaanxi Province under Grant No. 2024BSHYDZZ014, the ENN’s Hydrogen Boron Fusion Research Fund under 2025ENNHB01-013, the China Postdoctoral Science Foundation under Grant No. 2024M762569, the Shaanxi Fundamental Science Research Project for Mathematics and Physics under Grant No. 25JSQ025, the Fundamental Research Funds for the Central Universities under Grant No. xzy012025080, the Shaanxi Key R\&D Program Project No. 2024PT-ZCK-83, and Innovative Scientific Program of CNNC.

\bibliography{reference}

\end{document}